# Data Protection for Data Privacy-A South African Problem?

Venessa Darwin[1] and Mike Nkongolo[1]

[1]Department of Informatics, University of Pretoria, Pretoria, South Africa

Abstract:                                                      June 16, 2023

This study proposes a comprehensive framework for enhancing data security and privacy within organizations through data protection awareness. It employs a quantitative method and survey research strategy to assess the level of data protection awareness among employees of a public organization. Pre- and post-questionnaires, along with an explanatory poster and a video discussion, are utilized to evaluate participants' understanding of data protection. The study aims to identify misconceptions about data protection by comparing the results. Continuous promotion of data protection awareness is recommended, particularly in hybrid working environments, emphasizing the importance of understanding the eight conditions of POPIA (Protection of Personal Information Act). The findings of this study contribute to strengthening data protection practices and cultivating a culture of data privacy within organizations. Keywords: Cybercrime, Data privacy, Data protection, Information security, POPIA, network security

1. INTRODUCTION

The implementation of the Protection of Personal Information Act (POPIA) and the shift to remote work during the pandemic presented challenges that required organizations to invest resources in promoting data protection awareness. However, it is crucial for employees to fully understand their responsibilities in implementing the training they have received and complying with the Act. Employees directly involved in handling organizational data play a vital role in ensuring data protection compliance and implementing effective measures. Access to information and knowledge significantly impacts an employee's ability to effectively execute data protection awareness. Sayers proposes a seven-step process for raising awareness, including identifying the subject, generating employee interest, building skills and optimism, providing facilitation, using simulations to encourage action, and reinforcing the message [1]. Data privacy faces two major threats. The first threat is the expanding influence of the internet and social media, which amplify the capabilities of data protection and surveillance. This leads individuals to unintentionally or knowingly disclose private information. The second, more significant threat is the growing significance and dependence on information in decision-making processes. Policymakers prioritize the value of information, even if it comes at the cost of compromising privacy rights [2]. People are increasingly sharing their private information, both consciously and unconsciously. Additionally, there is a more significant and concerning threat related to the growing reliance on information for decision-making. Policymakers, as highlighted by Mason [3], prioritize the value of information, even if it involves compromising the privacy of individuals. Despite the Generalised Agreement on Accounting Principles (GAAP) not recognizing digital personal data as an asset on balance sheets, major technology companies such as Apple, Microsoft, Amazon, Google (Alphabet), and Facebook (Meta) have effectively turned data into a valuable asset [4], [4]. Recognizing the significance of data, its protection, and the potential risks associated with its misuse is of utmost importance. Organizations that collect and handle personal data bear the responsibility of safeguarding it through secure data management practices [3]. The value of data extends beyond organizations that offer products and services to consumers, as it has demonstrated considerable worth to cybercriminals [5], [6], [7].

*A. Research Objective*

To achieve the research objective, a quantitative survey methodology was utilized to gather valuable feedback. Additionally, in order to promote data protection awareness in various work settings, employees were provided with a poster (Figure 1) and a video containing insights on data protection. The secondary objectives outlined below aim to further enhance and strengthen the primary objective of this research study. The objective of the pre-questionnaire is to assess the current level of data protection awareness among employees after they have undergone training. The objectives of the post-questionnaire are to assess the employee's understanding of the POPIA conditions and their awareness of organizational policies and procedures related to data protection.

Additionally, it aims to investigate how employees apply their data protection knowledge in a hybrid working environment, considering different processing conditions. The diagram depicted in Figure 2 illustrates the implementation of data protection awareness by employees. The primary research question for this study focuses on how employees apply their data protection training to ensure compliance with POPIA regulations during the processing of personal information. The objective of this research, depicted in Figure 3, is to investigate the effective application of data protection awareness by employees in remote and on-site working environments, while also assessing their understanding of data protection following training.

Data privacy, although lacking a legal definition, plays a crucial role in preventing malicious use of data by cybercriminals [9], [10], [11]. Scholars in the field have attempted to classify privacy concerns and intrusion categories, such as the taxonomy proposed by Finn and Wright [12], which outlines seven types of privacy as presented in Table I. Alan Westin [13] extensively examines the conflict between privacy and surveillance in modern society, exploring the societal significance of privacy and its various functions. While data and information have distinct definitions, they are frequently used interchangeably.

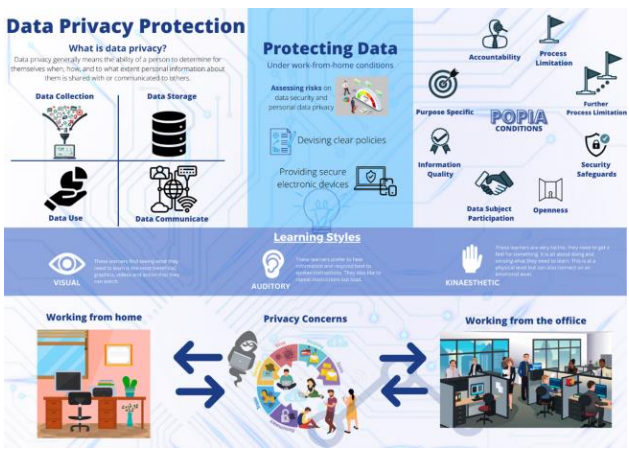

Figure 1. The data protection poster

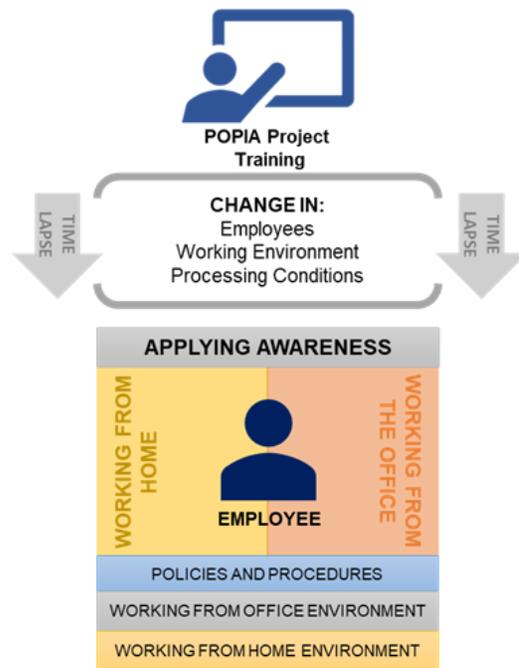

Figure 3. Research scope overview

The field of data protection can be classified into three primary domains: traditional data protection, data security, and data privacy (see Figure 4).

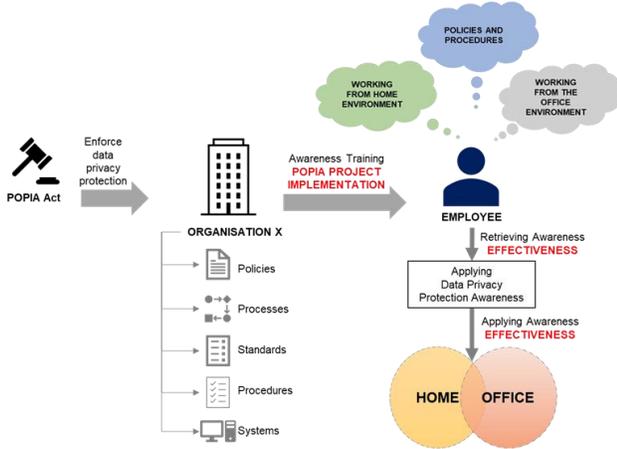

Figure 2. Assessment of the effective application of data protection awareness

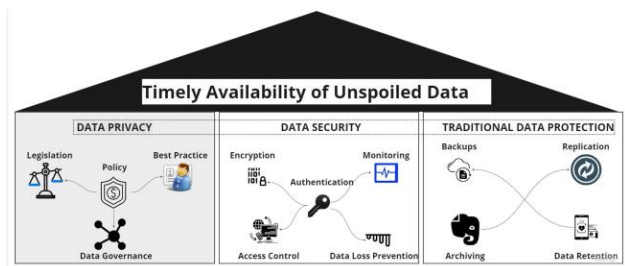

Figure 4. Data privacy, security, and protection

*B. The Privacy Concept*

According to the Cambridge English Dictionary, privacy is defined as the right of individuals to maintain the confidentiality of their personal affairs and relationships [8], encompassing personal space, information, and thoughts that are not meant to be shared.

Data, as defined by Baskarada [14], encompasses discrete pieces of information that may or may not be directly linked to a specific individual, possessing both quantitative and qualitative characteristics [15]. On the other hand, information is the result of combining various data attributes, providing deeper knowledge and understanding of a particular subject [14], [15]. Although the definitions of data and information privacy vary, they share a common focus on the management of personal information, with slight variations, as noted by Belanger et al. [16] and Rossi [15], leading to diverse and numerous challenges regarding privacy.

*C. Data Protection Guidelines*

In this section, we delve into the data protection guidelines established by South African legislation. However, it is crucial to acknowledge that data protection awareness and implementation should extend beyond mere compliance with legal requirements. As Doctorow [17] highlights, technological advancements and the internet present ongoing possibilities for the exploitation of information and data, even in the presence of regulatory oversight. Jordaan [18] has raised concerns regarding privacy, specifically addressing issues related to the collection of personal information, biases in data management, unauthorized access, and the potential misuse of sensitive data. Despite the existence of information protection legislation, certain government entities are exempt from data protection measures [19].

The exemption mentioned above highlights the challenge of striking a balance between the human right to privacy and the protection provided by governments to their citizens [20]. Furthermore, South Africa has witnessed a significant increase in cyber-attacks, with nearly 60% of the population utilizing various online devices [21], [22]. Cybercriminals focus their efforts on critical systems in order to infiltrate organizations and access valuable data and information [23]. Thus, it is crucial to establish robust data protection measures to mitigate these risks.

In South Africa, the annual cost of cybercrime victimization for the public amounts to approximately R2.2 billion, positioning the country as the second-highest in Africa for cybercrime incidents [24]. Cyber-attacks pose not only financial risks but also political and military threats, emphasizing the importance of data protection [23]. South Africa faces significant cyber security challenges, as reflected in its ranking as the fifth most affected country globally by cybercriminal activities (Figure 5). This ranking is based on the ratio of cybercrime victims per 1 million internet users, reported to the Federal Bureau of Investigation (FBI) from 2021 to 2022 [25], [26], [27]. Data protection awareness has become increasingly crucial in South Africa following cyberattacks such as the Dis-Chem and TransUnion breaches.

TABLE I. Privacy types (Finn and Wright [12])

| Number | Type of Privacy |
|---|---|
| 1 | A privacy related to a person A privacy related to an entity |
| 2 | Privacy related to behaviours and actions |
| 3 | Privacy related to communication |
| 4 | Privacy related to data and images |
| 5 | Privacy related to thoughts and feelings |
| 6 | Privacy related to space and location |
| 7 | Privacy related to associations Privacy of groups |

These attacks resulted in the compromise of millions of personal records, highlighting the need to safeguard the confidential information of stakeholders, partners, employees, and customers. The hacker group N4aughtysecTU was responsible for orchestrating these breaches. [1] . Moreover, numerous South African organizations, including Experian, City of Johannesburg, Ster-Kinekor, FNB, Standard Bank, ABSA, Telkom, the Department of Justice, Transnet, Uber Hack, and Momentum Metropolitan, have fallen victim to cybercrimes, as documented in a study by Akinbowale et al. [28].

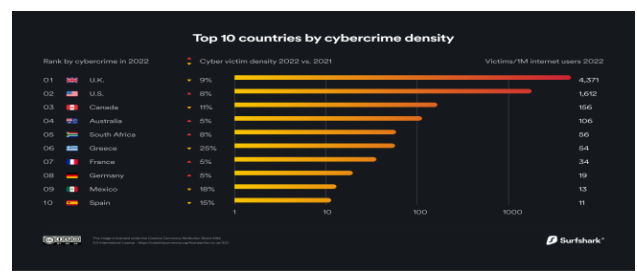

Figure 5. Top ten countries by cybercrime density

---

[1] https://www.itweb.co.za/content/PmxVE7KEABOqQY85

These incidents underscore the urgent need for enhanced data protection and security measures in South Africa. According to a 2017 study that examined South African data, the most prevalent types of cyber-attacks in the country are data exposure and financial theft [28]. The incident involving the former Public Protector Advocate Thuli Madonsela in South Africa highlights the importance of data protection, as she experienced a WhatsApp scam resulting in financial loss.[1] Madonsela shared her ordeal on Twitter, revealing how her friend's WhatsApp account was hacked, and the hacker demanded money over several months. This incident serves as a clear example of the need to prioritize data security in South Africa. Van Niekerk (2017) conducted an analysis of 54 cyber incidents, focusing on the potential impacts of various types of attacks, including Denial of Services (DoS), defacement, data corruption, system penetration, men in the middle, financial theft, and data exposure. The study revealed that data exposure and data corruption accounted for 45% of the overall impact of cyber incidents. Additionally, it was found that 54% of cyber attacks targeted state or political entities in South Africa. The prioritization of data protection measures varies among organizations [29]. Netshakhuma and Nkholedzeni [30] noted that public entities in South Africa generate critical information and data required for ensuring compliance with the country's laws and regulations.

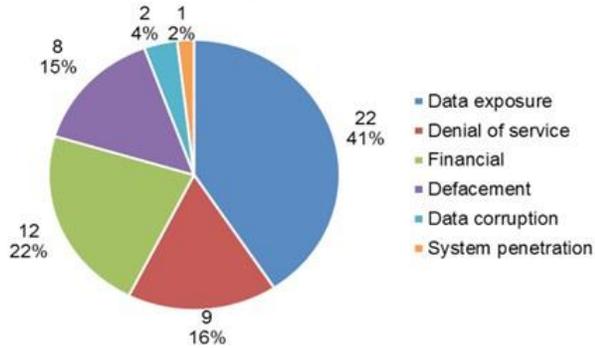

Figure 6. Analysis of cyber incidents in South Africa (Niekerk [21])

Ensuring data protection and compliance with POPIA regulations is crucial for the South African organization as they collect and process customer information. These organization must implement several measures to align with the requirements of POPIA. Figure 7 presents a comparison between POPIA compliance and the factors that impact the new normal of South Africans working and learning in a hybrid environment.

Figure 7. Employee's adherence and application of personal information

The comparison involves assessing the following aspects: employees' knowledge gaps regarding POPIA compliance, the necessity of compliance in any working environment, and adherence to organizational policies, standards, and procedures for data protection.

*D. The South African Data Privacy Legislation*

The South African Department of Justice and Constitutional Development introduced the POPIA on 14 December 2018, allowing organizations a one-year grace period to comply with its requirements [31]. To prepare for enforcement, many organizations established policies and codes of conduct as guidelines for information protection. The POPIA was officially promulgated by the president on 1 July 2020, aiming to ensure ethical management of data protection. During the COVID-19 pandemic, South Africa enforced a National State of Disaster from 15 March 2020 to 4 April 2022, which resulted in a nationwide lockdown with varying alert levels. This period coincided with the implementation of the POPIA by many organizations in South Africa, as they adapted to new work arrangements such as office-based, home-based, or hybrid models [32], [31]. The introduction of these new work arrangements played a crucial role in promoting data protection awareness and ensuring its effective enforcement throughout organizations. The primary goal of the POPIA is to ensure the security of sensitive data handled by South African organizations and to regulate the transfer of personal information across national borders. The Act outlines eight conditions (Table II) that require data protection awareness and compliance at all levels within organizations. In order to safeguard personal information, several countries, including South Africa with the POPIA and the General Data Protection Regulation (GDPR) as highlighted by David Banisar [27], have implemented measures to protect personal data. The reason behind this progress is that the recognition of data protection as a crucial component of human rights and privacy has led to its widespread adoption. Figure 8 demonstrates the status of data protection implementation across countries in 2023, including those that have successfully enacted it, those in the process of enacting it, and those lacking any data protection initiatives, such as the Democratic Republic of the Congo.

---

[1] https://www.thesouthafrican.com/news/advocate-thuli-madonselawhatsapp-scam-8-june-2023/

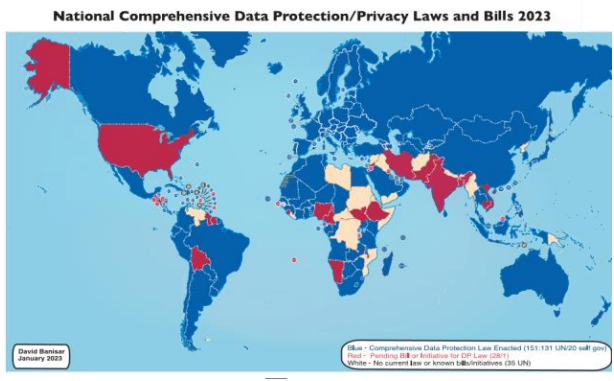

Figure 8. National comprehensive data protection/privacy laws and bills (Banisar [27])

To ensure the protection of people's privacy, a multitude of data protection legislations have been implemented worldwide. Banisar [27] emphasizes the progress made on an international scale from December 2016 to August 2023. Figure 8 clearly illustrates the evolving global landscape of data protection. South Africa has implemented various laws and regulations, including the Protection of Personal Information Act (POPIA), to safeguard privacy. These measures are rooted in the Constitution, which guarantees the right to privacy and explicitly mentions the right to communication privacy. The South African Electronic Communications and Transactions Act (ECTA) regulates electronic communication and includes data protection principles for data collectors to follow. Additionally, the Cybercrimes Act of 2020 addresses cybercrime offenses and imposes penalties for unauthorized data access, website tampering, virus insertion, and fraudulent activities for financial gain. The Cybercrimes Act provides guidelines for addressing cybercrimes and malicious communications of private data, as outlined in Chapter 2 of the Act.

This study posits that South African organisation's employees have access to information and security policies that guide them in managing, processing, and protecting data in accordance with the protection guidelines established by these Acts.

## 2. RESEARCH METHODOLOGY

This research aims to propose a framework for assessing employees' awareness, responsibility, and accountability towards data security in the context of data protection. Specifically, it focuses on evaluating the effectiveness of training provided during the implementation of the Protection of Personal Information Act (POPIA). The study explores the level of employees' knowledge and application of data protection principles, considering the unique circumstances of remote work during the COVID-19 pandemic. It investigates challenges faced by employees in implementing data protection practices while working from home and examines the influence of organizational culture on promoting a data protection mindset. Additionally, the research examines the impact of the hybrid work environment on data protection practices and awareness among employees. The research seeks to make a valuable contribution by developing a framework for enhancing data protection awareness in a hybrid work environment, thus ensuring the security of employees' data. This study adopts a research methodology based on the research onion framework introduced by Saunders et al. [33]. The selection of research options aligns with the different layers of the research onion, as illustrated in Figure 9.

TABLE II. The POPIA conditions

| Number | Condition | Expectation |
|---|---|---|
| 1 | Accountability | The party responsible for ensuring the lawful processing |
| 2 | Processing Limitations | Retention and restriction of records |
| 3 | Purpose Specific | Retention and limitation of records |
| 4 | Further Processing Limitations | Align with intended purpose of data collection |
| 5 | Information Quality | The standard of information accuracy and reliability |
| 6 | Security Safeguards | Confidentiality and integrity to maintain data security |
| 7 | Openness | Document and inform the subject about the data collection process |
| 8 | Data Subject Participation | The correction of information should be transparent, and accessible |

These options, including the philosophical stance, approach to theory development, methodological choice, research strategy, and time horizon, are elaborated in subsequent subsections to provide a comprehensive understanding. By employing this methodological approach, the study aims to conduct a systematic and rigorous investigation into the research problem at hand.

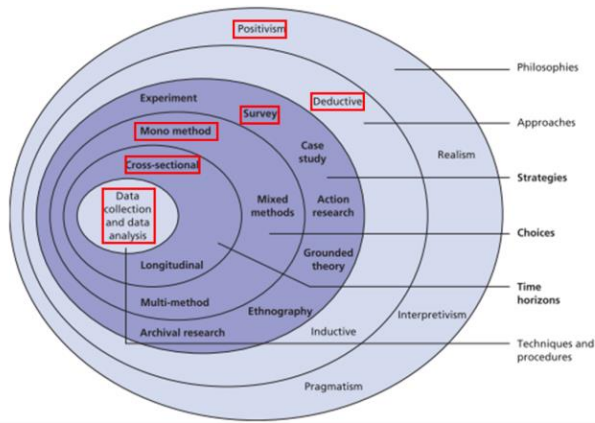

Figure 9. The research methodology (Saunders et al. [33])

*A. Research Philosophy*

Saunders et al. [33] proposed a research onion framework that includes various philosophical positions. The four main positions are positivism, interpretivism, realism, and pragmatism. Positivism emphasizes objective research with empirical and measurable data. Interpretivism acknowledges subjectivity and focuses on understanding the social and cultural context [33]. Realism combines empirical data and theoretical concepts to understand an objective reality. Pragmatism advocates for a flexible approach depending on the research problem [33]. These philosophical positions influence the research approach, methods, and data interpretation. In this study, a positivist research philosophy was adopted to ensure objectivity in data collection and interpretation. We aim to maintain minimal interaction with participants to remain independent. The positivist philosophy aligns with the belief that knowledge can be obtained through empirical observation and scientific methods. It is suitable for hypothesis testing and establishing causal relationships using empirical data.

*B. Research Approach*

This study proposes a deductive research approach based on the research onion model proposed by Saunders et al. [33]. The deductive involves the use of questionnaires to gather empirical data and compare different perspectives. It starts with a theory and formulate hypotheses or questions, which were tested through data collection. In contrast to the inductive approach that seeks to gain insights into the meaning attached to events, the deductive approach followed a structured process with controls in place to ensure the validity of the collected data. The study's outcomes can be analyzed to determine whether the hypotheses formulated for the research are confirmed or rejected.

*C. Research Choice*

The selected methodological approach for this study aligns with Saunders' research onion model [33] and is categorized as a mono quantitative method. This approach differs from mixed and multi-method research, as it solely focuses on either qualitative or quantitative data collection and analysis techniques. In this particular study, a mono method quantitative approach is proposed to ensure objectivity and precision in capturing participants' responses. This approach enabled efficient and timely data collection.

*D. Data Collection and Data Analysis*

Figure 10 illustrates the data collection process proposed in this study. Two surveys are designed using Microsoft Forms to gather the necessary data. The first survey focused on assessing participants' understanding of data protection, specifically in relation to the training received on the POPIA. The training for participants in this study can be conducted remotely through the use of animated videos, followed by a test focusing on a specific condition related to the POPIA. After being exposed to a data protection-related poster artifact, participants can complete a second survey. Additionally, a video can be provided to further explain the content of the poster after the initial survey completion. The surveys can consist of True or False questions, Likert Rating scale questions, and questions allowing for multiple answers. The questions in the second survey can be primarily based on the main themes presented in the poster. The study proposes a comparative analysis to assess the level of data protection awareness among participants before and after their exposure to the poster and the accompanying explanatory video. Table III provides an overview of the research methodology proposed in this study.

*E. Data Analysis Method*

Future works can aim to ascertain the level of data protection awareness among participants by employing a quantitative approach to data analysis. By utilizing numerical data, researchers can process and extract meaningful insights, which can be presented using charts, graphs, and statistics. The survey questionnaire can be designed to collect both opinions and factual data, with predefined answers to specific questions generating the latter. Following data collection, an exploratory and descriptive analysis can be conducted to gain a deeper understanding of the collected data. Researchers can then design and carry out research experiments to further investigate the level of data protection awareness among participants. Ultimately, these future works can contribute to a comprehensive conclusion regarding the participants' awareness of data protection.

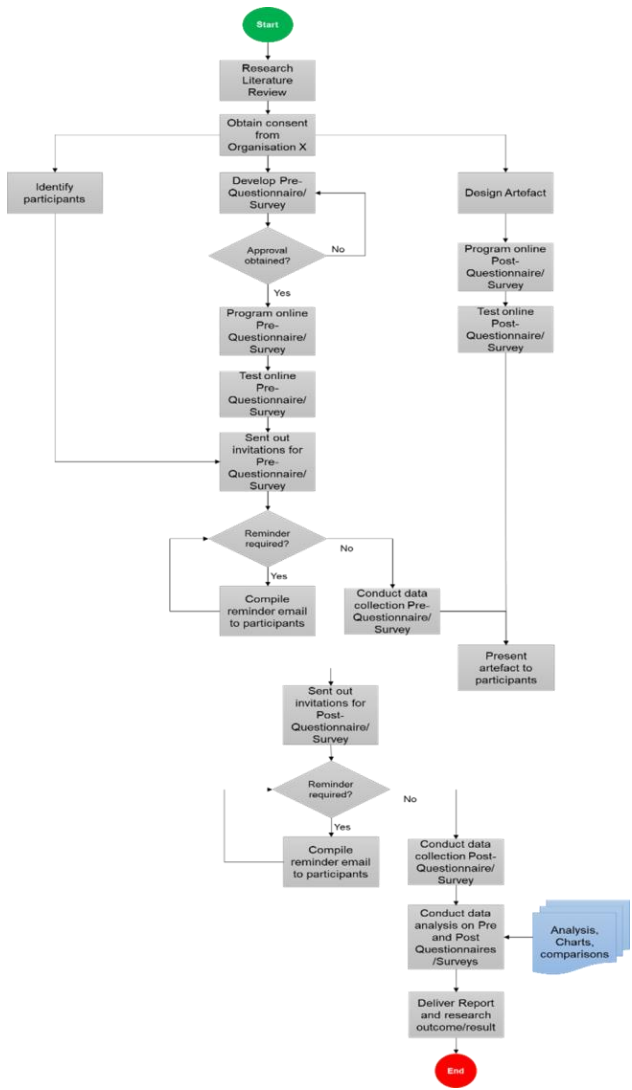

Figure 10. Outline of the research process

Furthermore, several limitations were identified within this study. Firstly, there might be challenges associated with the research sample and the selection process, potentially introducing bias into the study. Secondly, the sample size of the study may not be extensive enough to detect significant relationships within the data, despite efforts made to obtain a sizable sample. These limitations should be considered when interpreting the findings of this research. One significant limitation identified in this study is the lack of previous research on the POPIA. This knowledge gap hinders the ability to build upon existing literature and make meaningful comparisons. To address this limitation, further research should focus on assessing data protection awareness specifically within the context of the POPIA. This will contribute to a better understanding of data protection practices and aid in the development of effective strategies to safeguard data.

## 4. CONCLUSION

Assessing the level of data protection awareness among employees is a challenging task for organizations. While achieving optimal effectiveness in this area may be difficult, identifying gaps and deficiencies serves as a valuable starting point. To attain organizational objectives, the presence of policies, standards, procedures, and business processes is crucial, with employees playing a significant role in their implementation. With this in mind, this research investigated the impact of data protection awareness, taking into account individual employee preferences and circumstances that influence productivity. Sustained and ongoing refresher training and awareness initiatives are essential to ensure consistent and improved implementation of data protection measures.

## 3. LIMITATIONS

While numerous organizations have implemented initiatives to enhance data protection awareness, this study focuses exclusively on South African organizations and their workforce. Consequently, the framework presented in this research is limited to the context of South Africa.

TABLE III. Research methodology summary

|   | Research onion concepts | Research study selection |
|---|---|---|
| 1 | Philosophy | Positivist |
| 2 | Theory development | Deductivist |
| 3 | Methodological choice | Mono quantitative |
| 4 | Strategy | Survey |
| 5 | Time horizon | Cross-sectional |

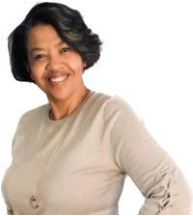

**Venessa Darwin** is an accomplished IT (Information Technology) professional with a track record of 27 years in utilizing technology to drive organizational growth, performance, and profitability. Her expertise includes implementing management frameworks to facilitate efficient and effective organizational functions that contribute to the overall strategy. She obtained her BCom in Business Informatics from the School of Economics and Management Sciences at the University of South Africa (UNISA), and completed her BCom Honours in Business Informatics from the School of Science, Engineering, and Technology at the same University. Venessa is currently pursuing a Master's degree in IT at the University of Pretoria under the supervision of Mr. Mike Nkongolo Wa Nkongolo. She specializes in the application of computational thinking and machine learning techniques to address data privacy concerns.

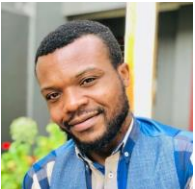

**Mike Nkongolo Wa Nkongolo** is a Lecturer in the Department of Informatics at the University of Pretoria. He holds a BSc in IT from Universite Protestante de Lubumbashi in the Democratic Republic of the Congo (DRC). He furthered his studies in South Africa, earning a Hdip, BSc Honours, and master's degrees in computer science from the University of the Witwatersrand at the School of Computer Science and Applied Mathematics. He has recently completed his Ph.D. thesis in IT at the University of Pretoria in the department of Informatics. Prior to his academic career, Mike worked in the telecommunication industry as a Systems Engineer, Technical Support Specialist, and Data Analyst. He is actively involved in the academic community, serving as a reviewer for IEEE Transactions on Education and as a programme committee member (Computer Science Track) at the South African Computer Scientists & Information Technologists Conference. Mike's research interests span several areas, including Network Security (Cryptography, Deep Packet Inspection, Intrusion Detection/Prevention, Data Protection), Artificial Intelligence, Machine Learning, Information Retrieval, Natural Language Processing, and Game Theory. Currently, he is supervising Venessa Darwin for her Master's in Information Technology (MIT) at the University of Pretoria in the department of Informatics.